\documentclass[%
 aps,
 amsmath,amssymb,
 reprint,%
floatfix,
]{revtex4-1}

\usepackage{amsfonts}
\usepackage{amsmath}
\usepackage{amssymb}
\usepackage{graphicx}
\usepackage{mathrsfs}
\usepackage{color}
\usepackage{bbold}
\usepackage{CJK}
\usepackage{times}
\usepackage{hyperref}
\usepackage{mathtools}

\usepackage{graphicx}
\usepackage{dcolumn}
\usepackage{bm}

\usepackage[utf8]{inputenc}
\usepackage[T1]{fontenc}
\usepackage{mathptmx}
\usepackage{color}

\begin{document}

\title[]{Ultra linear magnetic flux-to-voltage conversion in superconducting quantum interference proximity transistors}

\author{Giorgio De Simoni}
\email{giorgio.desimoni@sns.it}
\affiliation{NEST, Istituto Nanoscienze-CNR and Scuola Normale Superiore, I-56127 Pisa, Italy}
\author{Francesco Giazotto}
\email{francesco.giazotto@sns.it}
\affiliation{NEST, Istituto Nanoscienze-CNR and Scuola Normale Superiore, I-56127 Pisa, Italy}

\preprint{AIP/123-QED}

\begin{abstract}
Superconducting interferometers are quantum devices able to transduce a magnetic flux into an electrical output with excellent sensitivity, integrability and power consumption. Yet, their voltage response is intrinsically non-linear, a limitation which is conventionally circumvented through the introduction of compensation inductances or by the construction of complex device arrays. Here we propose an intrinsically-linear flux-to-voltage mesoscopic transducer, called bi-SQUIPT, based on the superconducting quantum interference proximity transistor as fundamental building block. The bi-SQUIPT provides a voltage-noise spectral density as low as $\sim10^{-16}$ V/Hz$^{1/2}$ and, more interestingly, under a proper operation parameter selection, exhibits a  spur-free dynamic range as large as $\sim60$ dB,  a value on par with that obtained with state-of-the-art SQUID-based linear flux-to-voltage superconducting transducers. Furthermore, thanks to its peculiar measurement configuration, the bi-SQUIPT is tolerant to imperfections and non-idealities in general. For the above reasons, we believe that the bi-SQUIPT could provide a relevant step-beyond in the field of low-dissipation and low-noise current amplification with a special emphasis on applications in cryogenic quantum electronics.

\end{abstract}

\maketitle

\section{Introduction}
\label{sec:Introduction}

Superconducting interferometers are quantum devices able to transduce a magnetic flux into an electrical output. Thanks to their high sensitivity, simple integrability and low heat dissipation, they are the key building block to implement ultra-sensitive cryogenic magnetometers and inductively-coupled current amplifiers \cite{Clarke2004,Barone1982,Kleiner2004,Martinez-Perez2017,Granata2016,Fagaly2015}. Moreover, they can be exploited as stand-alone devices or they can be included into larger systems, \textit{e. g.}, for signal processing applications \cite{Kornev2017a}. The direct-current (DC) SQUID (which stands for Superconducting Quantum Interference Device) is the almost ubiquitous implementation of superconducting interferometers. It consists of a pair of Josephson junctions (JJ) \cite{Josephson1962} closed on a superconducting ring, whose switching current is modulated by the magnetic flux $\phi$ with a periodicity equal to the magnetic flux quantum $\phi_0=h/2e$ \cite{Clarke2004,Barone1982,Doll1961,Deaver1961}. In practical applications, DC-SQUIDs are operated in the dissipative regime, \textit{i. e.}, they are current biased slightly-above their switching point while the voltage-drop across the loop ($V$) is measured \textit{vs.} the variation of $\phi$. In this configuration, excellent sensitivity can be achieved, but poor performances are observed in terms of linearity of the flux-to-voltage conversion. This behaviour is due to the quasi-sinusoidal $V vs. \phi$ SQUID characteristics, and although routinely it is exploited advantageously for magnetometry, it constitutes a drawback for the realization of current amplifiers. For the latter case, response linearity is a major requirement, which is generally improved through external reaction loops \cite{Clarke2004,Kleiner2004,Fagaly2015} or through the construction of arrays of SQUIDs \cite{Kornev2017a,Kornev2017}, at the cost of a worsening of the operation bandwidth and device integrability. A solution to these restrictions has arisen from the introduction of multi-loop superconducting interferometers, the main types of which are bi-SQUIDs \cite{Sharafiev2012,Kornev2014,Kornev2009,Kornev2011a, Sharafiev2012,DeSimoni2021bisquid} and D-SQUIDs \cite{Soloviev2019,Drung2013}. 

In bi-SQUIDs, a third JJ, closed on a larger superconducting ring, is placed in parallel to a smaller DC-SQUID to compensate for its non-linear response. Although the significant improvement of  their response linearity, Nb bi-SQUIDs showed a far from ideal performance, due to a high inductance and to the large area of the JJs \cite{Sharafiev2012,Kornev2020}. This made again necessary to include bi-SQUIDs in matrices of several devices.  At the cost of a severe design complexity and a large footprint area, bi-SQUIDs arrays are nonetheless extremely effective for low-noise magnetic field linear conversion \cite{Kornev2017,Kornev2017a,Kornev2011a,Kornev2020}.

The D-SQUIDs \cite{Soloviev2019,Drung2013} is the parallel of two identical DC-SQUIDs whose voltage drop is measured differentially and in which a constant additional magnetic flux component is added in one of the two rings. Such devices provide an effective linearization of the voltage response, and allow to obtain a high common-mode rejection ratio \cite{Drung2013}, a feature particularly useful when long wiring is required. Furthermore, it was shown that the effects of background magnetic fields and signal fluctuations related to temperature drifts are suppressed due to the differential reading \cite{Drung2013}. Nonetheless, it does not appear at present day that D-SQUIDs have yet found exploitation in real-world applications.

An alternative approach to conventional superconducting magnetometry is provided by superconducting quantum interference proximity transistors or SQUIPTs \cite{Giazotto2010,Giazotto2011,Ronzani2014,DAmbrosio2015,Virtanen2018,Jabdaraghi2017,Virtanen2016,Paolucci2022,Strambini2016,Ligato2022,Ligato2021}. These devices include a superconducting ring closed on a Josephson junction, generally consisting of a normal-metal constriction or constituted by a superconducting wire \cite{Virtanen2016,Ronzani2017,Ligato2021,Ligato2022}. Although the vast majority of JJs exploit the conventional superconducting/insulating/superconducting (SIS) scheme, the Josephson effect \cite{Likharev1979} is established also in weak-links based on a superconductor (SS$'$S)\cite{Vijay2010,Levenson-Falk2013}, or on a normal electron gas (N, either hosted by a semiconductor \cite{Giazotto2004,Carillo2006,Giazotto2011} or by a metal \cite{Savin2004a,DeSimoniACSALM2022,DeSimoniACSNANO2019}) enclosed between a pair of superconducting contacts (SNS). The latter support a non-dissipative current as a consequence of the formation of the Andreev bound states in the N region \cite{Pannetier2000a,Belzig1999,McMillan1968a}. 
This mechanism is called superconducting \textit{proximity} effect \cite{Pannetier2000a}, and has the remarkable consequence that a minigap opens in the weak-link density of states (DOS), having an amplitude that depends on the phase difference $\varphi$ of the superconducting order parameter at the ends of the junction. The latter, is then bound to the magnetic flux threading the loop by the fluxoid quantization relation. The information on the variation of $\phi$ is gained by measuring the current ($I$) \textit{vs.} voltage ($V$) characteristics of a superconducting or normal-metal tunnel electrode \cite{DAmbrosio2015} coupled to the weak-link by a thin insulating barrier through which a current can flow. The latter is, in fact, determined by the phase-dependent DOS in the weak-link. So far, SQUIPTs built with a few different geometries and material combinations \cite{Giazotto2010,Strambini2016,Meschke2011,DAmbrosio2015,Ronzani2017}, demonstrated to achieve a flux sensitivity a few orders of magnitude above their theoretical flux-noise boundary of $\sim $ n$\phi_0 / \sqrt{Hz}$ \cite{Ronzani2014,Ronzani2017,Giazotto2011}. Furthermore, they allow for ultra-low power dissipation thanks to a tunnel resistance that can be easily set in the $10^4-10^6$ $\Omega$ range, a value to be compared with that of a few $\Omega$ of the shunt resistor conventionally exploited in SQUID devices. Finally, thanks to the reduced inductance and to the low tunnel-junction capacitance, their operation bandwidth can be extended, in principle, up to the tens of GHz range \cite{Giazotto2013}. Yet, similarly to SQUIDs, SQUIPTs have a highly non-linear voltage-to-flux response, making so far their exploitation not convenient for cryogenic amplification.

\begin{figure}[t!]
  \includegraphics[width=0.97\columnwidth]{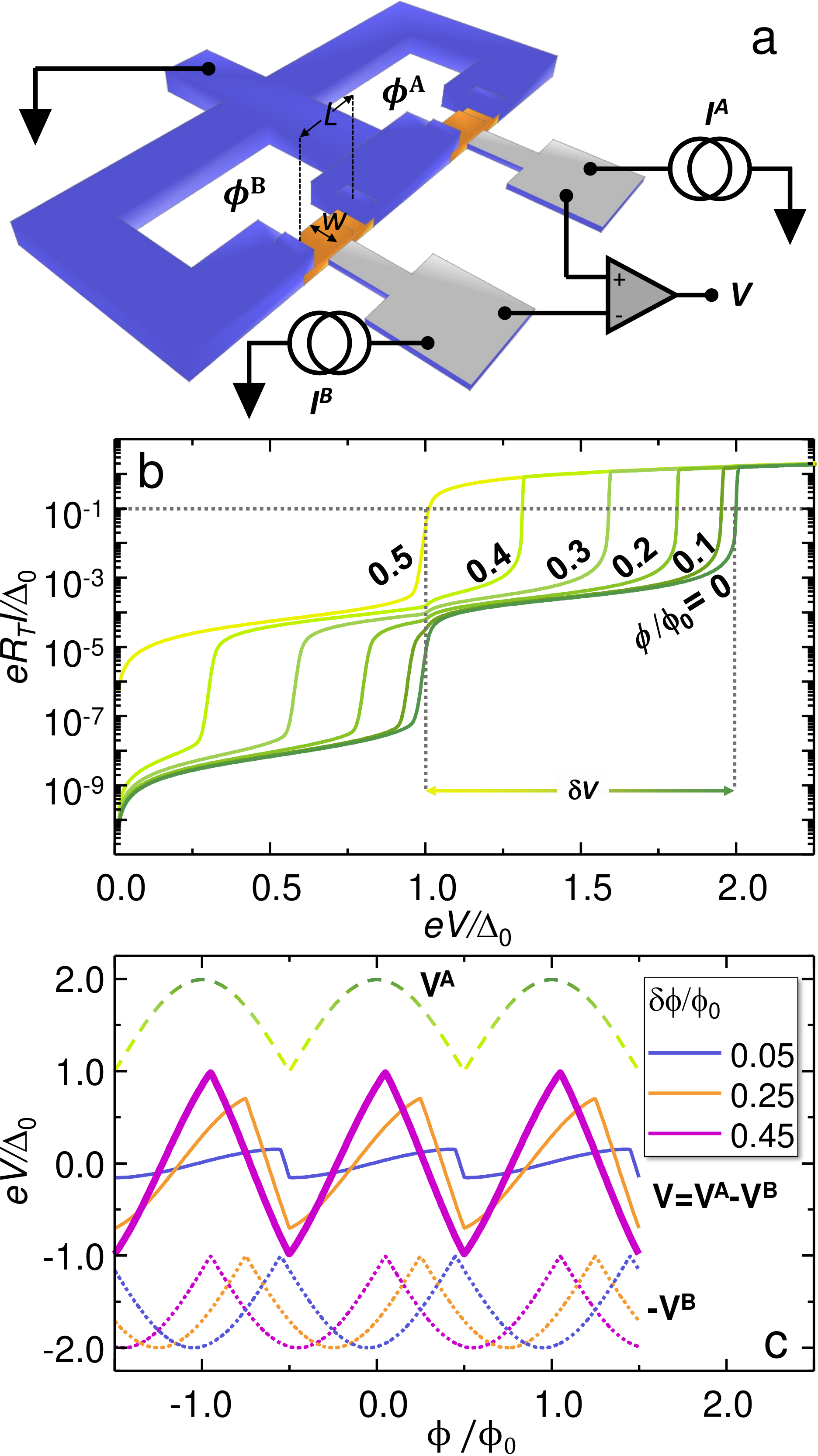}
    \caption{\textbf{Double superconducting quantum interference proximity transistor - The bi-SQUIPT} \textbf{a}: Two superconducting rings ($A$ and $B$, colored in blue), having the same area $A^A=A^B$, are closed on  $L$-long normal-metal weak-links  (orange). A magnetic flux $\phi_A=\phi$ and $\phi_B=\phi + \delta \phi$ is threaded to loop $A$ and $B$, respectively. A $w$-wide probe (made of the same superconductor, blue) is tunnel-coupled to each weak-link through a thin insulating layer (grey), and is exploited to inject  quasi-particle currents $I^A$ and $I^B$ and to measure the differential voltage drop $V=V^A-V^B$ with respect to the common ground terminal \textbf{b}: Normalized current ($I^{A,B}$) \textit{vs.} normalized voltage ($V^{A,B}$) of each tunnel junction calculated at a temperature $T=0.01 T_C$, where $T_C$ is the critical temperature of the superconductor, and for a Dynes parameter $\gamma=10^{-4} \Delta_0$, where $\Delta_0$ is the zero-temperature superconducting energy gap. $R_T$ is the tunnel resistance, and $e$ the electron charge. The curves are calculated for different values of $\phi^{A,B}$. At constant current bias (see dashed grey line) $V^{A,B}$ \textit{vs.} $\phi^{A,B}$ spans an interval $\delta V$. \textbf{c}: $V^A$ (dashed green curve), $-V^B$ (dotted lines), and $V=V^A-V^B$ (solid lines) as a function of $\phi$ for selected values of the $\delta \phi$. These curves were obtained by setting the same temperature and Dynes parameter of panel b.}
  \label{fig:fig1}
\end{figure}

Here we propose and theoretically analyze the bi-SQUIPT \cite{patent}, i.e., a linear-response three-terminal double-loop superconducting flux-to-voltage transducer which exploits the SQUIPT as a unit cell. We show that such a device, while retaining all the aforementioned SQUIPT characteristics, can compete in terms of linearity with SQUIDs and bi-SQUIDs arrays. Furthermore, similarly to D-SQUIDs, the bi-SQUIPT is measured in a differential configuration which promises a performance that is robust against temperature and magnetic flux drifts. This makes the bi-SQUIPT a promising candidate for the realization of ultra-low dissipation and ultra-low noise cryogenic current amplifiers to be integrated in quantum electronics setups.

\section{The bi-SQUIPT}
\label{sec:bisquipt}
A schematic diagram of a bi-SQUIPT is shown in Fig. \ref{fig:fig1}a. It consists of a three-terminal double-loop superconducting interferometer. The two loops have ideally the same area. The first terminal is the ground contact common to the loops, which are represented in blue. Each loop is closed on a $L$-long normal-metal (N) weak-link (orange in Fig. \ref{fig:fig1}a). The two $w$-wide probes are placed in the middle of the weak-links and provide the other two terminals of the device. They are tunnel-coupled to the N regions through a thin insulating layer (gray in Fig. \ref{fig:fig1}a). Although the tunnel probes can be either realized with a normal metal or a superconductor, in the following we will assume them to be made of the same superconducting material S of the loops. This choice is motivated by the ease of nano-fabrication of high quality SIN junctions due to the possibility of making the tunnel barriers (I) through a controlled oxidation of the surface of superconductors such as Al, which is the conventional material of choice to realize the probe electrodes as well as the loops of  superconducting interferometers. The case in which the probe is made of a normal metal will be discussed separately in section \ref{sec:Nprobe}, in order to allow for a comparison of these two different approaches.

We refer now to the quasiparticle current flowing through the first and second tunnel junctions as $I^A$ and $I^B$. They can be written as function of the voltage drops $V^{A,B}$ as \cite{Giazotto2011}
\begin{equation}
\label{eq:IV}
\begin{split}
I^{A,B}(V^{A,B})=\frac{1}{ewR^{A,B}}\times
\\
\int_{\frac{L-w}{2}}^{\frac{L+w}{2}}dx\int_{-\infty}^{\infty}d\varepsilon \mathcal{D}_w^{A,B}(x,\varepsilon,\varphi^{A,B},T) \times
\\
\mathcal{D}_{p}^{A,B}(\tilde{\varepsilon}^{A,B},T)F(\varepsilon,\tilde{\varepsilon}^{A,B},T),
\end{split}
\end{equation} 
where $\varepsilon$ is the energy relative to the chemical potential in the superconductor, $\tilde{\varepsilon}^{a,B}=\varepsilon-eV^{A,B}$, $F(\varepsilon,\tilde{\varepsilon}^{A,B})=[f_0(\tilde{\varepsilon}^{A,B})-f_0(\varepsilon)]$, $f_0(\varepsilon)$ is the Fermi-Dirac energy distribution function, $R^{A,B}$ are the normal-state tunneling resistance of the probing junctions, $x \in[0,L]$ is the spatial coordinate along the N weak-links, and indexes A or B refer to the weak-link A and B, respectively.

$\mathcal{D}_w^{A,B}$ and $\mathcal{D}_p^{A,B}(\varepsilon,T)=|\Re[(\varepsilon+i\Gamma)/\sqrt{(\varepsilon+i\Gamma)^2-\Delta^2(T)}]|$ are the DOSs at electronic temperature $T$ of the S probes  and of the N weak-link, respectively, where $\Delta(T) \simeq \Delta_0 \tanh(1.74 \sqrt{(T/T_c)^{-1} - 1}])$ accounts for the temperature evolution of the superconducting order parameter $\Delta$ with respect to its zero-temperature value $\Delta_0$, $T_C$ is the critical temperature of the superconductor, and $\Gamma$ is the Dynes parameter accounting for the phenomenological DOS broadening. In the short-junction limit, which we adopt for simplicity in the present analysis and because in this limit proximity effect is maximized in the weak-links, it holds \cite{Heikkila2002, Virtanen2018}:
\begin{equation}
\label{equation:DosW}
  \begin{split}
  \mathcal{D}_w^{A,B}(x,\varphi^{A,B},\varepsilon,T)
  =\\
  \Re\sqrt{\frac{(\varepsilon+i\Gamma)^2}{(\varepsilon+i\Gamma)^2-\Delta^2(T)\cos^2\frac{\varphi^{A,B} }{2}}}\times\\
  \cosh\Bigl(
  \frac{2x-L}{L}\text{arcosh}\sqrt{\frac{(\varepsilon+i\Gamma)^2-\Delta^2(T)\text{cos}^2\frac{\varphi^{A,B} }{2}}{(\varepsilon+i\Gamma)^2-\Delta^2(T)}}\Bigr),
  \end{split}
\end{equation}
where for sake of clarity we assume $\Gamma$ to have the same value for both the probes and the weak-links. Equation \ref{equation:DosW} implies that $\mathcal{D}_w^{A,B}$ exhibits a minigap $\varepsilon_g^w (\varphi^{A,B} ,T)=\Delta(T)|\text{cos}(\varphi^{A,B}/2)|$, which turns out to be  spatially constant along the weak-links. In particular, $\varepsilon_g^{A,B}=\Delta$ for $\varphi^{A,B}=0$ and vanishes at $\varphi^{A,B} =\pi$. These extrema correspond to a quasiparticle spectrum in the weak-links which is made equivalent to that of a superconducting material or to that of a normal conductor just by changing the phase-drops across the weak-links. The latter are linked to to the magnetic flux threading the loops.

A properly designed bi-SQUIPT should comply the following conditions:
\begin{equation}
2\pi I_C L^{loop} \lesssim \phi_0,
\label{condi}
\end{equation}
\begin{equation}
L_k^{loop} \ll L_k^w
\label{condii}
\end{equation}
where $I_C$ and $L_k^{w}$ are the critical supercurrent and the kinetic inductance of each weak-link, respectively, and $L^{loop}$ is the total loop inductance corresponding to the sum of the kinetic ($L_k^{loop}$) and geometric ($L_g^{loop}$) contributions. In particular, condition (\ref{condi}) avoids magnetic hysteresis, while condition (\ref{condii}) ensures that the phase difference $\varphi^{A,B}$ drops entirely at SNS junction ends, thereby allowing for a full modulation of their DOSs. Under the above assumptions the fluxoid quantization can be simply expressed as
\begin{equation}
    \varphi^{A,B}=2\pi \phi^{A,B} / \phi_0.
\end{equation}
It was shown that both conditions can be easily fulfilled by a proper choice of materials and device geometry\cite{leSueur2008,Meschke2011}.

The normalized $I \,vs \,V^{A,B}$ characteristics at $T=0.01 T_C$ of the tunnel junctions is shown in Fig. \ref{fig:fig1}b, for selected value of $\phi^{A,B}$ between 0 and $0.5 \phi_0$, and $\Gamma=10^{-4} \Delta_0$. The $I(V^{A,B})$s reflect the evolution of the DOS of the N weak-link with the flux, and evolve from that of S/I/S junction, at $\phi^{A,B}=0$, to that of a S/I/normal-metal, for $\phi^{A,B}=0.5 \phi_0$. 
In the intermediate flux range, at constant current bias, the shrinking of the gap results in a reduction of the voltage drop, which is minimum (minigap completely closed) at $\phi^{A,B}=0.5 \phi_0$, and maximum at $\phi^{A,B}=0$. 
Therefore, $V^{A,B}$ spans a range $\delta V$ that can be, at most, as large as $\Delta_0$. For instance, such a value is reached by setting $I^{A,B}=0.1 I_0 e R_T^{A,B}/ \Delta_0$ (see the gray dotted line in Fig. \ref{fig:fig1}b), with $\delta V \sim [\Delta_0,2 \Delta_0]$. The $V^A vs. \phi^A $ characteristics is shown in Fig. \ref{fig:fig1}c (green dashed line), for $T=0.1 T_C$ and $I^{A}=0.1 e R_T^{A,B}/ \Delta_0$. Such a curve,  which is calculate by solving Eq. (\ref{eq:IV}) at fixed bias current as a function of $\phi^A$ shows that each loop of the bi-SQUIPT behaves as a flux-to-voltage transducer, as expected. 

The bi-SQUIPT response $V=V^A-V^B$ is obtained by subtraction of voltage responses of its parts: for a perfectly symmetrical device (same areas and $R^A=R^B=R_T$) $V=0$ for each flux $\phi^A=\phi^B=\phi$. However, by introducing an additional flux component $\delta \phi$ such that $\phi^B=\phi+\delta \phi$, $V$ is not identical to 0 any more. Fig. \ref{fig:fig1}c shows $-V^B(\phi)$ for selected values of $\delta \phi$ (dotted lines), which are equivalent to shifted and mirrored clones of $V^A$. Therefore, due to the symmetry and periodicity of the SQUIPT response, the differential measurement of junctions A and B allows to subtract from the SQUIPT voltage \textit{vs.} flux characteristics its mirrored image. This leads to a partial compensation of the nonlinear terms of the flux-to-voltage transfer function, as shown in Fig. \ref{fig:fig1}c, where $V$ is plotted $vs.$  flux for selected values of $\delta \phi$ (solid lines). In particular, for $\delta \phi= 0.45 \phi_0$, $V$ has quasi-triangular shape and spans a voltage range of about $2 \Delta_0/e$, that is twice larger than the single SQUIPT response. On the latter point, we emphasize that $\delta V$ is a function of $\delta \phi$, that can be exploited as knob to tune the magnification factor of the bi-SQUIPT response. These characteristics make the bi-SQUIPT an interesting and promising platform to implement superconducting inductively-coupled linear-response current preamplifiers.

The amplitudes of bias currents are the other main knobs influencing the bi-SQUIPT behavior. Figure \ref{fig:fig2}a, reports $V(\phi)$ for selected values of $I^A=I^B=I$ at $\delta \phi=0.45 \phi_0$. For the lowest biases, the shape of the voltage-to-flux characteristics retains the peculiar triangular shape up to about $0.45 \Delta_0/e R_T$. Above such a threshold the response becomes progressively distorted, due to a rounding of the  curves at their extrema. The almost-linear behavior is instead retained around the zeroes for a much larger range of current (up to $I\sim 0.1  \Delta_0/e R_T$).

\section{Evaluation of the linearity of the flux-to-voltage response}
\label{sec:linearity}
\begin{figure}[t!]
  \includegraphics[width=1.0\columnwidth]{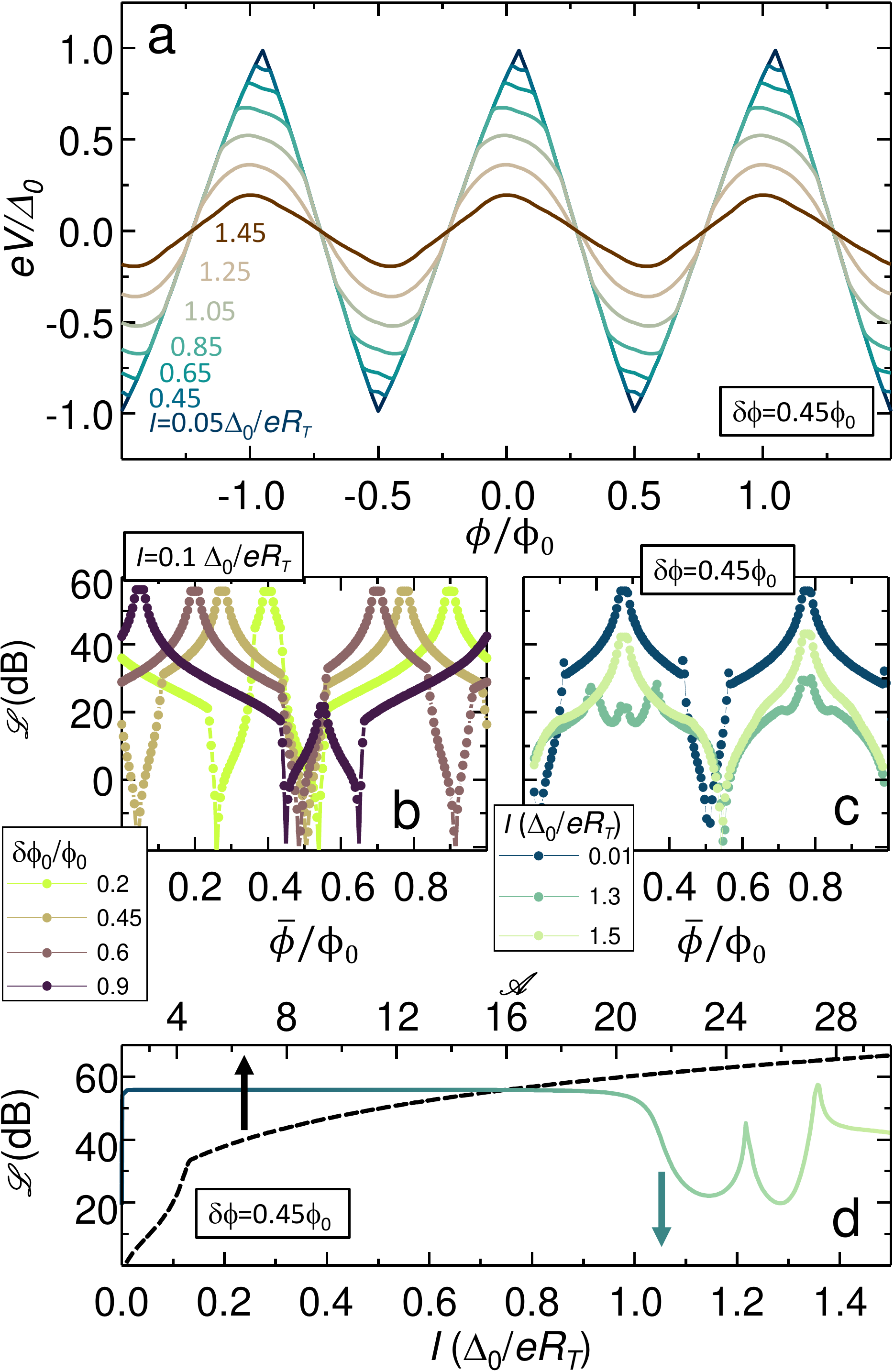}
    \caption{\textbf{Linearity of the bi-SQUIPT output} \textbf{a}: Voltage output $V$ of an ideal bi-SQUIPT as a function of the flux $\phi$ for $\delta \phi=0.45 \phi_0$, where $\phi_0$ is the magnetic flux quantum, for selected values of the current bias $I^A=I^B=I$, at $T=0.01 T_C$ and for $\Gamma=10^{-4} \Delta_0$ \textbf{b,c}: Spurious-free dynamic range (SFDR) $\mathcal{L}$ of an ideal bi-SQUIPT as a function of the flux working point $\overline{\phi}$  at constant current bias $I=0.1 \Delta_0=/e R_T$ for selected values of $\delta \phi$ (panel b), and at constant flux shift $\delta \phi=0.45 \phi_0$ for selected current bias $I$ (panel c). Curves were obtained by setting $T=0.01 T_C$ and $\Gamma=10^{-4} \Delta_0$. \textbf{d}: Evolution of $\mathcal{L}$ \textit{vs.}  flux modulation amplitude $\mathcal{A}/\phi_0$ (black dashed line, top axis) and \textit{vs.}  bias current $I$ (aquamarine solid line, bottom axis).}
  \label{fig:fig2}
\end{figure}

The linearity can be quantified by the conventional figure of merit of the total spur-free dynamic range (SFDR) $\mathcal{L}_{dB} (\overline{\phi})=-20 log(\frac{A_M(\overline{\phi})}{A_1(\overline{\phi})})$, defined as the ratio of the amplitude of the carrier wave (maximum signal component) $A_1$ to the amplitude of the next largest noise or harmonic distortion component $A_M$ at the output of the bi-SQUIPT. We calculated $\mathcal{L}$  at a fixed flux working point $\overline{\phi}$ through the Fourier transform $A_0+\sum_{n} A_n sin(\omega_n t )$ of the device voltage response $V(\phi(t))$ to a periodic sinusoidal modulation of the magnetic flux of amplitude $\phi_0/\mathcal{A}$, such that $\phi(t)=\overline{\phi}+\frac{\phi_0}{\mathcal{A}} sin(\omega_1 t)$, where $t$ is the time coordinate. 
This method is valid, in principle, in the limit $\omega_1 \ll 1/\tau$, where $\tau$ is the response time of the system, determined by the minimum of the Thouless frequency of the weak-link $f_{Th}=E_{Th} / 2 \pi\hbar$, the frequency of the gap of superconductors $f_{\Delta_0}=\Delta_0/2\pi\hbar$, and  the cut-off frequencies  $f_{RL}=R_T/[2\pi (L^{loop}+L_k^w)]$, $f_{LC}=1/[2\pi \sqrt{(L^{loop}+L_k^w)C}]$ and $f_{RC}=1/(2\pi R_T C)$ (where $C$ is the tunnel junction capacitance) accounting for the inductive/resistive, inductive/capacitive and resistive/capacitive time-scale, respectively. In the short junction-limit and for probe electrode and loops both made of Al $f_{Th} \sim f_{\Delta_0} \sim 100$ GHz. Through a proper device design $(L^{loop}+L_k^w)$ and $C$ can be set respectively in the pH and  fF range, leading to $f_{RL}\sim 10^{15}$ Hz, $f_{LC}\sim 10^{12}$ Hz and $f_{RC}\sim 10^{10}$ Hz. Therefore, the resistive/capacitive time-scale set the upper limit for $\omega_1$ in the $\lesssim 10$ GHz range, a value which is about from two to  three orders of magnitude larger than commercially available SQUID-based current amplifiers, and substantially on par with bi-SQUID based devices \cite{Prokopenko2013}.

Figure \ref{fig:fig2}b shows $\mathcal{L} vs. \overline{\phi}$ curves for selected values of $\delta \phi$, for $T=0.01 T_C$, $I=0.1 \Delta_0/e R_T$, and $\mathcal{A}=16$. $\mathcal{L}$ spans a large interval of values as a function of $\overline{\phi}$ ranging from a minimum of $\sim-20$ dB (\textit{i. e.}, the response is strongly non linear) in the surrounding of the extrema of the  flux-to-voltage characteristics, to a maximum value as large as $\sim 60$ dB around the zeroes. Such performance is equivalent to that of SNS mesoscopic bi-SQUIDs \cite{DeSimoni2021bisquid} and close to that of arrays of bi-SQUIDs. Interestingly, by changing the value of $\delta \phi$, the maximum value of $\mathcal{L}$ shifts accordingly. This behavior can be conveniently exploited to tune the bi-SQUIPT response to the needed flux working point $\overline{\phi}$ during device operation. On the other hand, the current bias knob marginally impacts the device linearity for values lower than $\sim \Delta_0/e R_T$. For larger values, as shown in Fig. \ref{fig:fig2}c for selected values of $I$ ($\mathcal{A}=16$ and $\delta \phi=0.45 \phi_0$), the $\mathcal{L} vs. \overline{\phi}$ curves are strongly modified. $\mathcal{L}(\delta \phi=0.45)$ is plotted $vs. I$ in Fig. \ref{fig:fig2}d (green solid line) for $\mathcal{A}=16$. It is, indeed, flat up to $I\sim \Delta_0/e R_T$, a value above which the linearity evolves non monotonically. The maximum SFDR  value is obtained for $I\sim 1.4\Delta_0/e R_T$, nonetheless we identify the best operating range for $I$ in the $[0.01,1]$ interval, in which the linearity performance is expected to be very robust to inaccuracies   in the current bias. In Fig. \ref{fig:fig2}d the evolution of the SFDR as a function of the flux modulation amplitude is also shown (black dashed line). For modulation amplitudes  larger than $\sim \phi_0/4$, the linearity drops, due to the residual non-linearity in the neighbourhood of the extrema of the $V vs. \phi$ characteristics. By increasing $A$, $\mathcal{L}$ monotonically grows, exceeding the value of $40 $ dB for $A\sim 5$, and approaching asymptotically the value of $\sim 70$ dB.

\section{Robustness of the linearity performance to device imperfections}
\label{sec:imperfections}
\begin{figure}[ht]
  \includegraphics[width=1\columnwidth]{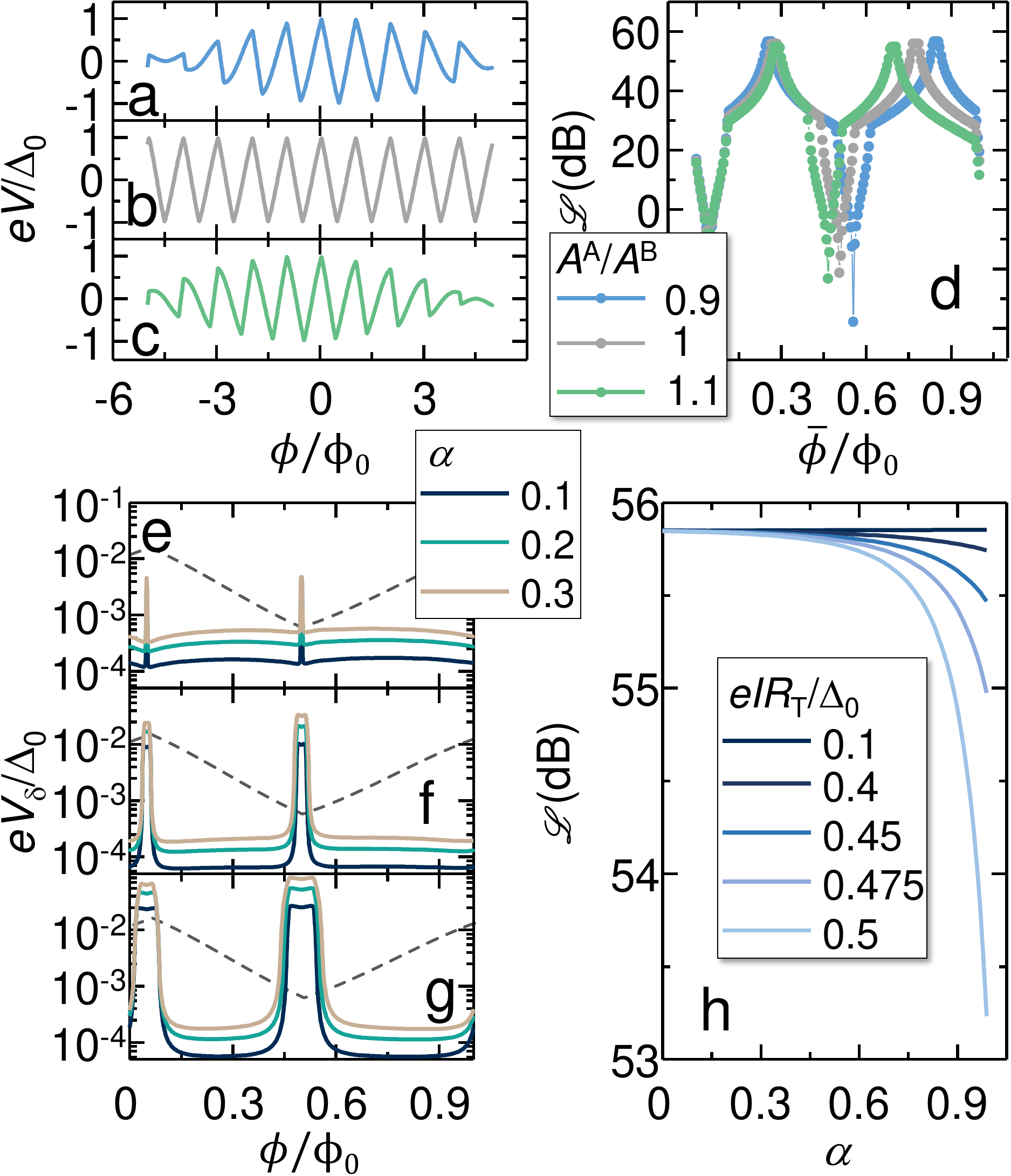}
    \caption{\textbf{Robustness of bi-SQUIPT performance to construction asymmetries} \textbf{a,b,c}: Asymmetric bi-SQUIPT output $V$as function $\phi$ for $\delta \phi=0.45 \phi_0$ and $I=0.1 \Delta_0/e R_T$, for selected values of the ratio between loop areas $A^{A}/A^{B}$, at $T=0.01 T_C$ and for $\Gamma=10^{-4} \Delta_0$ \textbf{d}: Spurious-free dynamic range (SFDR) $\mathcal{L}$ of the asymmetric bi-SQUIPT as a function of $\overline{\phi}$. Curves were obtained for the same parameters of panels a,b,c. \textbf{e,f,g}: Deviation $V_\delta$ of the output voltage of bi-SQUIPT with asymmetric tunnel resistances ($R^A\neq R^B$) from the ideal case ($R^A=R^B=R_T$) for selected values of the asymmetry parameter $\alpha=\frac{2(R^A-R^B)}{R^A+R^B}$. Curves were obtained assuming $A^A=A^B$, and using the same parameters choice of panels a,b,c and d. \textbf{h}: SFDR $\mathcal{L}$ \textit{vs.} $\alpha$ for selected values of $I$.}
  \label{fig:fig3}
\end{figure}

In this section we discuss the impact on the performance of bi-SQUPTs of unavoidable construction non-idealities  present on real devices.
In the first place, we address the effect of a small difference between the areas $A^A$ and $A^B$ of the superconducting loop $A$ and $B$, respectively. The effect of a deviation from perfect symmetry results in a discrepancy between the periodicity of the voltage-flux characteristics of the loops. This leads to a flux-dependent phase-shift between $ V^A $ and $ V^B $, \textit{i.e.}, to a beating between the two output voltages. As a result, the voltage-flux characteristic of the entire device results to be dramatically affected, as shown in Fig. \ref{fig:fig3}a, b and c for values of the ratio $ A^B / A^A $ equal to 0.9, 1 and 1.1, respectively, and for $\delta \phi=0.45 \phi_0$, $T=0.01 T_C$ and  $I=0.1 \Delta_0/e R_T$. We wish to  emphasize that such inaccuracy on device areas corresponds to a linear error on the loop side of the order a few hundreds of nanometers (for a loop area of the order of  1 $\mu$m$^2$), \textit{i.e.}, well above the conventional resolution achieved by modern lithographic techniques. Furthermore, despite the considerable distortion of the voltage-flux characteristics, if we limit the operation range in the $[0,\phi_0$] interval, the linearity performance is only marginally affected, as shown in Fig. \ref{fig:fig3}d for the same selected value of $A_B/A_A$ of Fig. \ref{fig:fig3}a,b,c. Indeed, due to the beating of $V^A(\phi)$ and $V^B(\phi)$ the maxima of $\mathcal{L}$ shifts when $R_A/R_B\neq 1$, but its value remains almost unchanged.

Similarly, as long as the device is operated in the low-current regime ($I \ll \Delta_0 / e R_T$), a slight discrepancy of the tunnel resistances of the two probes affects only marginally the device performance. In Fig. \ref{fig:fig3}e, f, and g  we plot the deviation $V_{\delta}=|V(\phi ; \alpha=0)-V(\phi; \alpha\neq 0)|$ of the bi-SQUIPT voltage output from the perfectly symmetric case ($R^A=R^B$, $\alpha =0$) vs magnetic flux for three values of the resistance asymmetry parameter $\alpha=\frac{2(R^A-R^B)}{R^A+R^B}$. $V_\delta$ is lower than $10^{-3} \Delta_0/e$ for almost every value of $\phi$, with the exception around the extrema of the $V\, vs.\, \phi$ characteristics, where the deviation reaches at most the value of $10^{-1} \Delta_0/e$, and has anyway negligible impact on the linearity of the device (we remind that the best linearity response is obtained around the zeroes of the $V vs. \phi$). Such a behavior can be straightforwardly understood by examining the shape of the tunnel junction $V(I)$ characteristics (see Fig. \ref{fig:fig1}b) around the working point $I=0.1\Delta_0/e R_T$. At such current bias, the junctions are essentially operated \textit{below-threshold}, \textit{i.e.} where the tunnel resistance has a reduced impact on the conductance, being the $V(I)$ mainly governed by the Dynes $\Gamma$ (accounting for the sub-gap conductance, whose impact is discussed in section \ref{sec:temperature}).  Indeed, as shown in Fig. \ref{fig:fig3}e, $\mathcal{L}$ is flat \textit{vs.} $\alpha$ for $I=0.1 \Delta_0/ e R_T$, while becomes increasingly larger for $\alpha \gtrsim 0.5$ when the current is higher than $\sim 0.4 \Delta_0/ e R_T$.

From a general point of view, we conclude this section by emphasizing the expected theoretical
robustness of the linearity performance of bi-SQUIPTs with respect to construction imperfections. This is especially true in light of the potentiality of modern fabrication techniques, which allow for a nanometer-scale resolution as well as for a very strict control of junction resistance. 

\section{Impact of the temperature and of the Dynes parameter $\Gamma$}
\label{sec:temperature}

\begin{figure}[ht]
  \includegraphics[width=1\columnwidth]{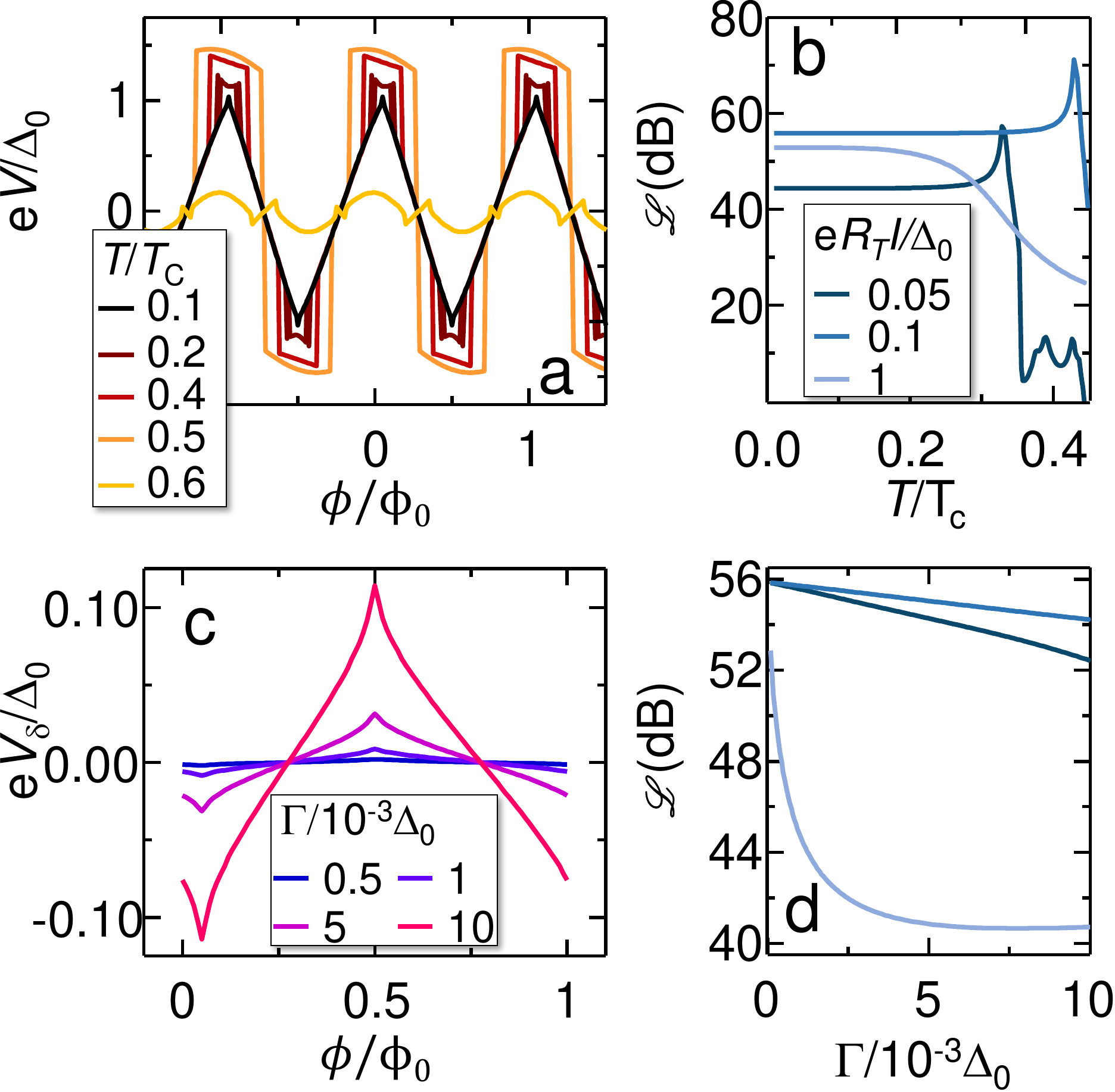}
    \caption{\textbf{Impact of the temperature and of the Dynes parameter  $\Gamma$} \textbf{a}: Normalized voltage output $V$ of an ideal bi-SQUIPT as a function of $\phi$ for different values of the normalized temperature $T/T_C$. Curves were obtained by setting $I=0.1 \Delta_0/e R_T$ and $\Gamma=10^{-4} \Delta_0$. \textbf{b}: SFDR $\mathcal{L}$ \textit{vs.} normalized temperature $T/T_C$ for selected values of the normalized current bias $I$. \textbf{c}: $V_{\delta}$ \textit{vs.} $\phi$ for some values of the Dynes parameter $\Gamma$. The curves were calculated by setting $I=0.1 \Delta_0/e R_T$ and $T=0.01 T_C$. \textbf{d}: SFDR $\mathcal{L}$ \textit{vs.} $\Gamma$ for selected values of the normalized current bias $I$. Color coding of the curves is the same of panel b.}
  \label{fig:fig4}
\end{figure}

Here, we investigate the temperature and the sub-gap conductance dependence of the bi-SQUIPT performance.

Figure \ref{fig:fig4}a shows the voltage output of an ideal bi-SQUIPT for selected values of the electronic temperature $T$ normalized on the critical temperature $T_C$, at $\delta \phi=0.45 \phi_0$ and for $I=0.1 \Delta_0 e R_T$. With this parameter setting, we note that the output characteristics remains essentially unchanged up to $T\sim 0.1 T_C$. At such a temperature, a significant deviation appears due to the appearance of the matching peaks in the quasiparticle current of the tunnel junctions, which reflects in a strong deformation of the $V-\phi$ curves around their maxima. This mainly results in the shrinking of the amplitude of the linear-response intervals in flux and, equivalently, in a temperature-dependent reduction of available dynamic range for flux modulation. 
These considerations are confirmed by the analysis of the SFDR evolution \textit{vs.}  temperature (calculated at $\overline{\phi}=0.45 \phi_0$), reported in \ref{fig:fig4}b for selected values of the bias current between $0.05 \Delta_0/ e R_T$ and $\Delta_0/ e R_T$. Although the linearity performance is determined by the current bias, the overall temperature evolution remains essentially the same up to $I=\Delta/e R_T$, with a plateau region (which reaches its largest value for $I=0.1 \Delta_0 /e R_T$) followed by a significant performance drop. In fact, the bi-SQUIPT can be fully exploited up $T\lesssim 0.5T_C$. Above this temperature the linearity interval is pinched-off and the typical triangular output is completely lost. If the device is based on low $T_C$ superconductors (such as Al), for some specific applications this temperature boundary might be considered a major limitation as it implies the use of $^3$He refrigerators. However, we emphasize that by simply switching to metallic superconductors with a higher critical temperature (\textit{e. g.} NbN or NbTiN) the operating range can be  extended above 4 K, a temperature that can routinely be obtained in $^4$He closed-cycle cryostats.     

Beside the temperature, the phenomenological broadening of the DOS, described by the Dynes $\Gamma$ parameter, is the other main mechanism responsible for an expected worsening of the bi-SQUIPT performance. $\Gamma$ is a parameter that empirically takes into account for the presence of quasiparticles  within the superconducting gap, and is  mainly determined by the quality of the superconducting material and of the tunnel junctions. Similarly to Fig. \ref{fig:fig3}e,f,g, Fig. \ref{fig:fig4}c shows the difference $V_\delta=V(\phi; \Gamma)- V(\phi; \Gamma=10^{-4} \Delta_0)$, for selected values in the range $[5 \times 10^{-4} \Delta_0,10^{-2} \Delta_0]$. $\Gamma$ was assumed to be the same for the loops and for the probes. 
In typical experiments, $\Gamma$ values as low as $\sim10^{-4} \Delta_0$ can be obtained in high-quality tunnel junctions. For this reason such a value was assumed to be the reference for the following discussion. Furthermore, the explored range is large enough to provide the best- and the worst-case scenarios for the bi-SQUIPTs performance. The plot of $V_\delta vs. \phi$ allows us to observe that the deviation is again relevant in the interval around the maxima, whose width increases with $\Gamma$, where a variation of about a $10 \%$ of the original total voltage output is reached at $\Gamma=10^{-2} \Delta_0$. 
Although such relatively large variation, this has a somewhat low impact on linearity because this is maximized at the nodes of the $V-\phi$ output characteristics. Indeed, by looking at the plot of the SFDR as a function of $\Gamma$ for selected current bias (shown in Fig. \ref{fig:fig4}d for $\delta \phi=0.45$ and $T=0.01 T_C$) we can notice an almost-linear, yet moderate, decrease in linearity as a function of $\Gamma$, for currents below $\sim0.1 \Delta_0=/e R_T$. At higher currents, the decrease is steeper, and the behavior is no more linear. This suggests (in the limit of low current bias) a substantial robustness of the proposed linearization scheme even at relatively large values of the Dynes parameter, and shows the actual bi-SQUIPT to be tolerant to a sub-ideal quality of the exploited superconducting material.

\section{bi-SQUIPTs based on normal-metal tunnel probes}
\label{sec:Nprobe}

\begin{figure}[ht]
  \includegraphics[width=1\columnwidth]{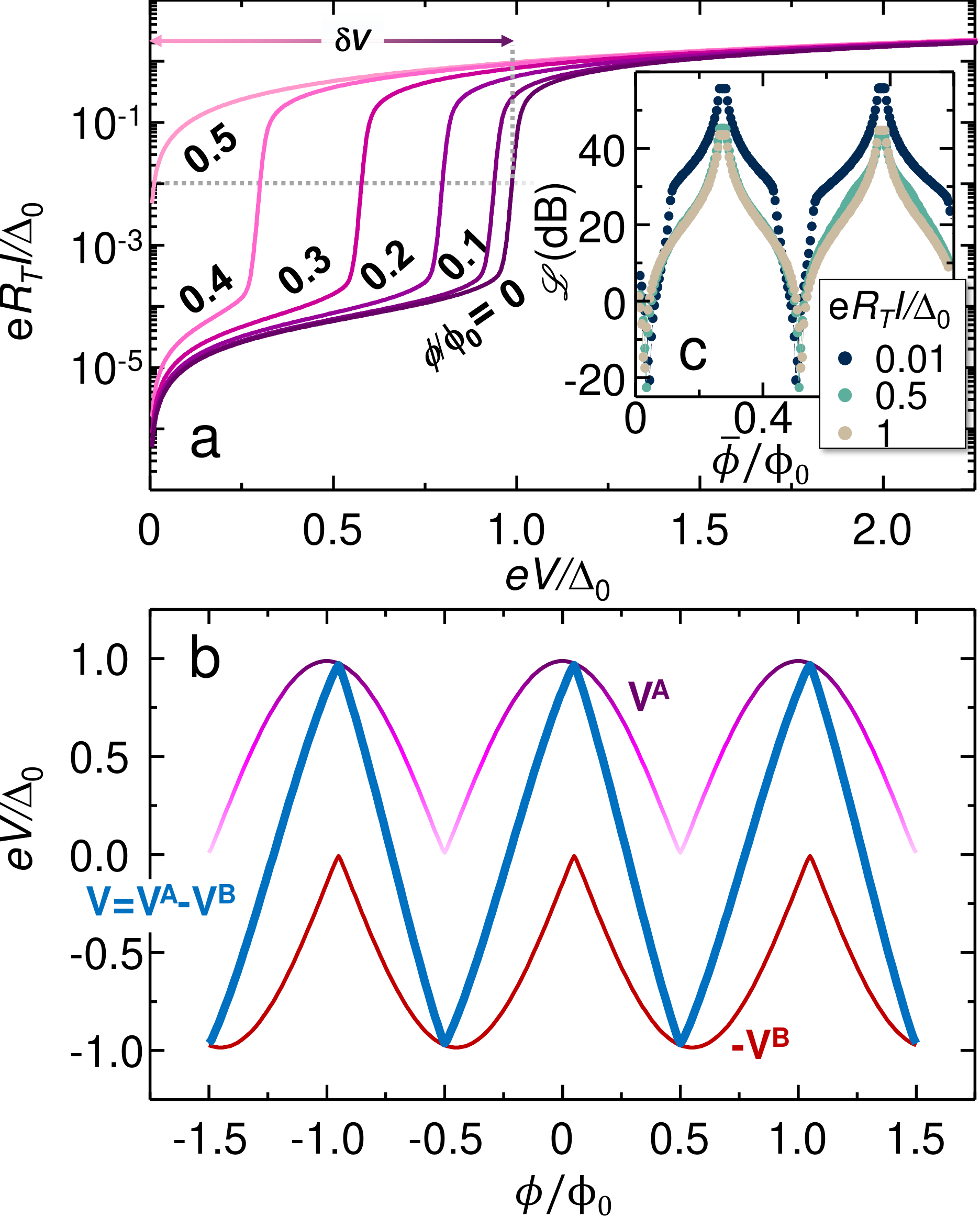}
    \caption{\textbf{bi-SQUIPT based on normal-metal tunnel probes} \textbf{a}: Normalized current ($I^{A,B}$) \textit{vs.} normalized voltage ($V^{A,B}$) of each tunnel junction of a bi-SQUIPT based on a N-type tunnel probe. Curves were calculated at $T=0.01 T_C$ and  $\Gamma=10^{-4} \Delta_0$ for different values of the magnetic flux $\phi^{A,B}$. Similarly to the S-type probe case, as function of the flux, at constant current bias (see the dashed grey guide for the eye) $V^{A,B}$ span an interval $\delta V$. \textbf{b}: $V^A$ (violet curve), $-V^B$ (red line), and $V=V^A-V^B$ (blue line) as a function of $\phi$ for $\delta \phi=0.45 \phi_0$. The curves were obtained by setting the same temperature and Dynes parameter as in panel a. \textbf{c }: SFDR $\mathcal{L}$ of a N-type probe bi-SQUIPT as a function of  $\overline{\phi}$ at constant flux shift $\delta \phi=0.45 \phi_0$ for selected current bias $I$. Curves were obtained by setting $T=0.01 T_C$ and $\Gamma=10^{-4} \Delta_0$. }
  \label{fig:fig5}
\end{figure}

In this section we discuss an alternative fabrication scheme for bi-SQUIPTs, in which the tunnel probes are made of a normal metal \cite{DAmbrosio2015}, instead of a superconductor. For conventional SQUIPT interferometers, the approach based on S-type tunnel probes is usually preferred because of a sharper response and improved noise performance, and since it allows an easy engineering of the tunnel junction insulating layer. The latter, in fact, can be easily fabricated via a controlled oxidation of an Al superconducting thin film, which provides a high-quality native insulating layer. Yet, the tunnel resistance can be tailored up to a large extent during the fabrication process either varying the oxygen pressure  or by changing the duration of the oxidation procedure. 
Such a convenient technique cannot be directly easily applied  to the most common normal metals exploited in mesoscopic device manufacturing such as, \textit{e. g.}, Cu unless using Al$_{0.98}$Mn$_{0.02}$ thin films, which are typically able to provide a native oxide \cite{DAmbrosio2015} suitable for the realization of high-quality tunnel junctions.  
Nonetheless, a N-type tunnel probe based approach allows a power dissipation about 2-times lower than the S-type one, thanks to the absence of the superconducting gap in the density of states of the tunnel probe. At the same time, the linearity performance is essentially the same of the S-type-probe scheme. For the above reasons, here we discuss such an alternative fabrication setup that might be conveniently exploited whenever power dissipation restriction is a  relevant issue.

The $I-V$ characteristics of the NIN tunnel junction can be calculated following the same procedure adopted for the SIN case. Differently from the latter, in Eq. (\ref{eq:IV}) the normalized superconducting DOS has to be replaced with that of normal metal ($\mathcal{D}_p^{A,B}=1$). 
The current-voltage characteristics of the single $A$ or $B$ SIN tunnel junction is shown in Fig. \ref{fig:fig5}a for selected values of the magnetic flux threading each loop and for $T=0.01 T_C$. 
We emphasize that, due to the presence in the junction of a single gapped material, the $I-V$ characteristics exhibit a conduction threshold at $V\simeq\Delta_0$. Similarly to the NIS case, $\epsilon_g^{A,B}$ is determined by the magnetic flux, and it is equal to $\Delta(T)$ when $\varphi^{A,B}(\phi)=0$, then decreasing down to zero when $\varphi^{A,B}(\phi)$ approaches $\pi$. When a flux $\phi$ is applied, the $I-V$ characteristic evolves from that of an NIN junction  to that of an SIN junction, i.e.,  with one \textit{gapped} and one \textit{gap-less} electrode. 

At constant current bias $I\sim0.01\Delta_0/eR_T$, the width interval $\delta V$ spanned by the voltage drop turns out to be maximum, and it is comprised between 0 and $\Delta(T)/e$. The $V^{A}$ \textit{vs.} $\phi$ characteristic is plotted (violet line) in Fig. \ref{fig:fig4}b, for $T=0.01 T_C$ and $I=0.01 \Delta_0/e R_T$. We emphasize that such characteristics resembles very closely that of a NIS junction, with the main exception of the spanned $V$ interval which is shifted to lower voltages by $\sim \Delta_0/e$. 
In this regard, it is worth to note that, with the same bias current, a bi-SQUIPT based on the N-type probe dissipates approximately 2-times less  than the one based on an S-type probe.

The similarity of the flux-to-voltage characteristics suggest a straightforward application of the linearization scheme demonstrated for the superconducting-probe case, \textit{i. e.}, we adopt as signal output the difference between the voltage drops $V=V^A-V^B$, when an additional flux $\delta \phi$ is added to the loop $B$ only. Figure \ref{fig:fig5}b shows the voltage probe on the junction $B$ ($-V^B$, red line) and the bi-SQUIPT output $V$ for, $\delta \phi =0.45 \phi_0$, $T=0.01 T_C$, and $I=0.01 \Delta_0/e R_T$. The $V-\phi$ curve can be hardly distinguished by the one obtained in the case of superconducting probes thereby suggesting that both the N-type- and the S-type-probe devices can be successfully exploited to realize bi-SQUIPT based linear-response amplifiers. This hypothesis is quantitatively confirmed by the calculation of the SFDR $\mathcal{L}$ of an N-type-probe bi-SQUIPT with the same technique used for the S-type-probe case. Figure \ref{fig:fig5}c shows the $\mathcal{L} vs. \overline{\phi}$ characteristics of an N-type probe bi-SQUIPT, at $T=0.01 T_C$ and with $\delta \phi=0.45 \phi_0$, for selected values of the current bias $I$. The maximum value ($\sim 60$ dB) of $\mathcal{L}$ is obtained for a bias current lower than $\sim0.01 \Delta_0 /e R_T$, but SFDR values al large as $\sim$45 dB can be obtained for currents up to $ \Delta_0/e R_T$.

We conclude this section by highlighting the substantial irrelevance of the tunnel probe construction technique for the purpose of maximizing the response linearity of the bi-SQUIPT. We believe that, thanks to the greater ease of fabrication of Al oxide tunnel layers, the approach based on superconducting probes is the one to be preferred for the majority of the cases. However, the improved performance in terms of dissipated power might make the use of a normal probe more attractive for some specific applications. In the next section, we shall complete the comparison between these two alternative approaches by discussing their characteristics in terms of noise spectral density.

\section{Calculation of the noise spectral density}
\label{sec:Noise}
\begin{figure}[ht]
  \includegraphics[width=1\columnwidth]{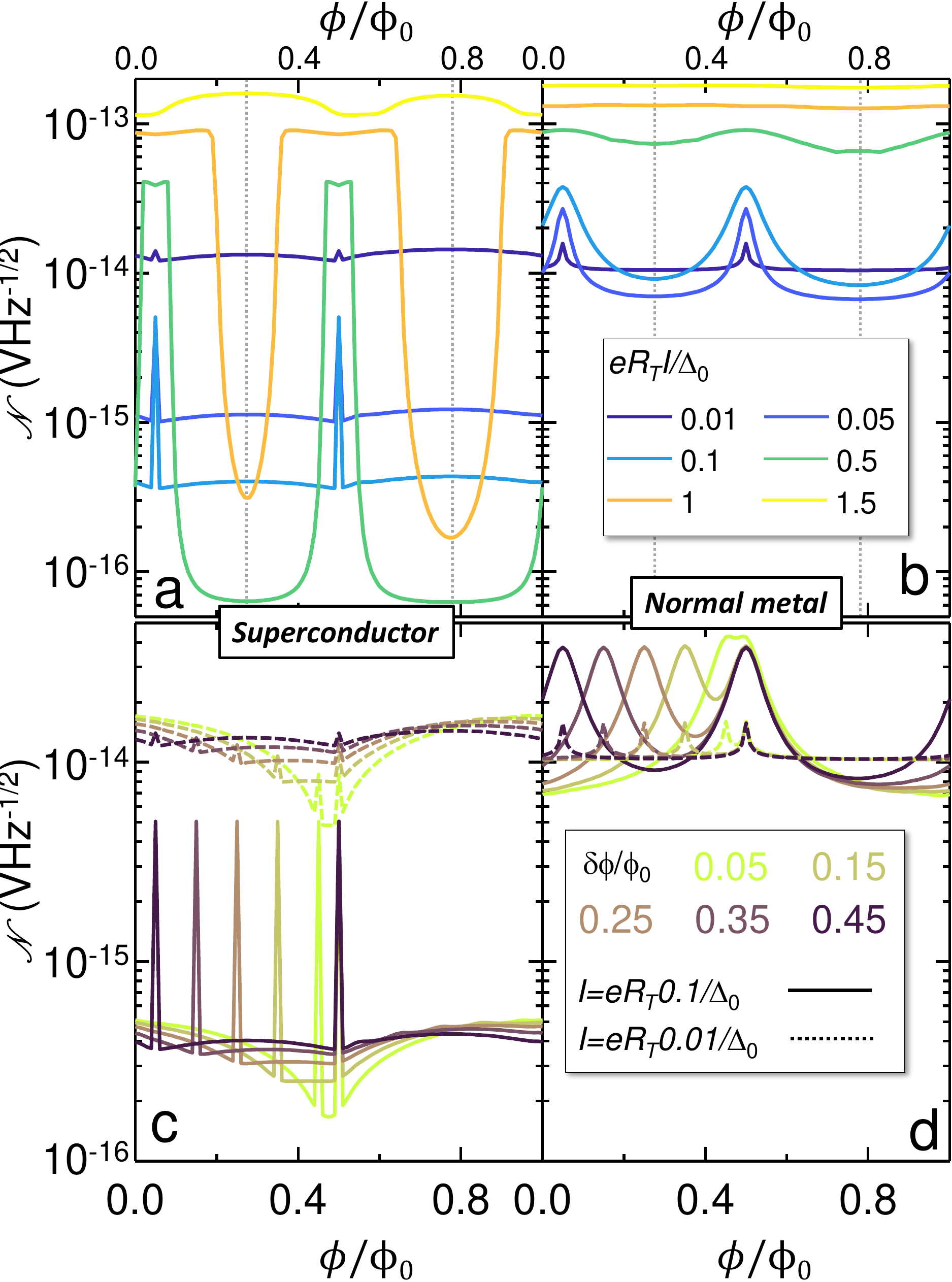}
    \caption{\textbf{Voltage-noise spectral density} \textbf{a,b}: Voltage-noise spectral density $\mathcal{N}$ as a function of $\phi$ of an ideal bi-SQUIPT based on a S-type probe (panel a) and on a N-type probe (panel b), for selected values of the current bias $I$.
    The calculation parameters are $0.01 T_C$,  $\delta \phi=0.45 \phi_0$, and $\Gamma=10^{-4} \Delta_0$. \textbf{c,d}: $\mathcal{N}$ \textit{vs.} $\phi$ characteristics of an ideal S-type probe (panel c)  and of  on a N-type probe (panel d) bi-SQUIPT, for selected values of the flux shift $\delta \phi$ and $I$. The other calculation parameters are the same of panels a and b.}
  \label{fig:fig6}
\end{figure}

In this last section we discuss the impact of quantum shot-noise on the behavior of bi-SQUIPTS. We note that  such a figure of merit is essentially  determined by the composition of the noise contribution from the the two tunnel junctions and, therefore, the total noise is merely the sum of that of two SQUIPTs. Indeed, this can be quantified through the quadrature sum of the voltage-noise spectral density of the tunnel junctions, which reads \cite{Paolucci2022}
\begin{equation}
\label{eq:noise}
\mathcal{N}=\sqrt{\sum_{A,B} \Bigl(\frac{dV^{A,B}}{dI^{A,B}}\Bigr)^2 2eI^{A,B} \coth \Bigl(\frac{eV^{A,B}}{2 k_B T}\Bigr)}.
\end{equation}
Figure \ref{fig:fig6}a and b shows the $\mathcal{N} vs. \phi $ characteristics of bi-SQUIPTS realized with S-type- and N-type probes, respectively, for different values of the current bias $I$ and for $\delta \phi=0.45 \phi_0$. The curves where obtained by assuming the devices to be made with a geometry and materials feasible by standard fabrication techniques. In particular, for the rings (and the tunnel probes, in S-type case) we set $\Delta_0 = \Delta_{Al}=200$ $\mu$eV (\textit{i. e.} the superconductor is aluminum), and we fixed the  tunnel resistance to a value of $R_T = 10^5$ $\Omega$. Finally, the temperature was set to $0.01 T_C\simeq 0.01 \Delta_{Al}/1.764 k_B\sim 13$ mK. In flux, $\mathcal{N}$ undergoes to large variations (up to $\sim3$ orders of magnitude, for $I=0.5 \Delta_0/e R_T$ in the S-type probe case), for both the normal-metal and the superconducting probes. In particular, we note that minimum noise is obtained in correspondence of the zeroes of the $V-\phi$ curves (highlighted by the dashed vertical gray lines in Fig. \ref{fig:fig6}a and b), \textit{i. e.}, where the linearity is maximized), while largest noise is expected at the extrema of the curves. Furthermore, for $I>0.01 \Delta_0/eR_T$ the noise spectral density obtained for the S-type probe device is in general lower than that of a N-type probe device, reaching the minimum value of $\sim 7 \times 10^{-17}$V/Hz$^{1/2}$ for $I=0.5 \Delta_0/e R_T$, a value consistent with that obtained for SQUIPT-based magnetometers \cite{Paolucci2022}. This observation suggests the S-type-probe bi-SQUIPT to be preferred when minimum noise is especially required. At the same time, N-type bi-SQUIPTs provide a feasible alternative approach to be used for reduced power dissipation applications, with an acceptable noise performance when operated at low current-bias. 

Besides, $\delta \phi$ has a noticeable impact on the shape of the voltage-noise spectral density. Indeed, by shifting the voltage-flux characteristics of one junction respect to the other, the noise-flux characteristics of the same junction is accordingly shifted as well. This behavior can be clearly noticed in Fig. \ref{fig:fig6}c and d (where $\mathcal{N}$ is plotted \textit{vs.} $\phi$ for selected values of $I$ and $\delta \phi$) by looking at the relative position of the peaks originating from the noise contribution of each junction, which is shrunk or enlarged depending on $\delta \phi$. We speculate that this characteristic might be exploited to minimize the voltage-noise spectral density at specific flux working-points.
\section{Conclusions}
\label{sec:Conclusions}

We have discussed a highly-linear three-terminal double-loop superconducting flux-to-voltage mesoscopic transducer called bi-SQUIPT, which is based on the superconducting quantum interference proximity transistor (SQUIPT). We have demonstrated the bi-SQUIPT to preserve the main SQUIPT characteristics such as the low-power dissipation, and the excellent noise performance. More interestingly, under proper operation parameters selection,  the bi-SQUIPT  exhibits highly-linear magnetic flux-to-voltage response characteristics. In this regards, we quantified the  spur-free dynamic range of the device response to be as large as $\sim60$ dB. Such a result is on par with that obtained with state-of-the-art SQUID-based linear flux-to-voltage transducers. Furthermore, thanks to its peculiar measurement configuration, the bi-SQUIPT is particularly robust against construction imperfections. For the above reasons,  we candidate the bi-SQUIPT as a possible key-tool for the future generation of  ultra-low dissipation and ultra-low noise cryogenic current amplifiers to be exploited in quantum electronics setups.

\section*{Acknowledgement}
The authors  acknowledge  the  EU’s  Horizon 2020 research and innovation program under Grant Agreement No. 800923 (SUPERTED) and No. 964398 (SUPERGATE) for partial financial support.



%

\end{document}